\documentstyle[sprocl]{article}

\begin{document}

\title{Tomographic map within the framework of star-product
quantization}

\author{Olga V. Man'ko and Vladimir I. Man'ko}
\address{ P. N. Lebedev
Physical Institute, Leninskii prospect 53, 119991 Moscow, Russia}
\author{Giuseppe Marmo}
\address{Dipartimento di Scienze Fisiche, Universit\`a 
``Federico II'' di Napoli and\\
Istituto Nazionale di Fisica Nucleare, Sezione di
Napoli\\ Complesso Universitario di Monte S'Angelo, via Cintia, 
80126 Napoli, Italy}

\maketitle\abstracts{ Tomograms introduced for the description of
quantum states in
terms of probability distributions are shown to be related to
a standard star-product quantization with appropriate kernels.
Examples of symplectic tomograms and spin tomograms are
presented.}

\section{Introduction}

Star-product quantization$\,$\cite{[1]}
provides us the a formulation of
quantum mechanics where quantum observables
are presented by functions on the phase
space instead of operators acting on some Hilbert space. 
The product of functions is not the point-wise product. 
This induced  product is nonlocal and
it is defined by some kernel. Recently the tomographic 
description of quantum
states has been introduced to associate states with  
standard
 probability distributions instead of density 
operators.\cite{Tombesi13,Tombesi13'}
The tomographic description was introduced also for 
spin.\cite{dod14,JETF}
Properties of tomographic maps and their relation
with Heisenberg--Weyl and SU(2)
groups were studied.\cite{21klimov}$^-$\cite{22klimov}
The relation of tomographic representation
to the star-product quantization was 
established.\cite{22klimov,JPAmarmotobepublished}
The aim of this contribution
is to review this approach$\,$\cite{22klimov,JPAmarmotobepublished} 
and to study properties
of the star-product of symbols of quantum observables. 
We show that
 the tomographic map can be discussed within 
the framework of the standard 
star-product quantization procedure.

\section{Star-product}

 In quantum mechanics, observables are described by 
operators acting on the Hilbert space of states. In order
to consider observables as functions on a phase 
space, we review first a general construction and
provide general relations and properties of a map from
operators onto functions without a concrete realization of the map. 
Given a Hilbert space $H$ and an operator $\hat A $
acting on this space, let us suppose that we have a set
of operators $\hat U({\bf x})$ and a $n$-dimensional
vector ${\bf x}=(x_1,x_2,\ldots,x_n)$. 
We construct the $c$-number function
$f_{\hat A}({\bf x})$ (we call it the symbol of operator $\hat A$ )
$$f_{\hat A}({\bf x})=\mbox{Tr}\left[\hat A\hat U({\bf x})\right].$$
Let us suppose that the relation has an inverse,
i.e., there exists a set of operators $\hat D({\bf x})$ 
such that
$$\hat A= \int f_{\hat A}({\bf x})\hat D({\bf x})~d{\bf x}. $$
In fact, we could consider maps of the form 
 
$$\hat A\rightarrow f_{\hat A}({\bf x})\qquad\mbox{and}\qquad 
f_{\hat A}({\bf x})\rightarrow\hat A,$$
and require them to be one the inverse of the other.

The most important property is the
existence of associative product (star-product) of the symbols 
induced
 by the product of operators. We introduce
the product (star-product) of two
functions $f_{\hat A}({\bf x})$ and $f_{\hat B}({\bf x})$
corresponding to two operators $\hat A$ and
$\hat B$ by the relations
$$f_{\hat A\hat B}({\bf x})=f_{\hat A}({\bf x})*
f_{\hat B} ({\bf x}):=\mbox{Tr}\left[\hat A\hat B\hat U({\bf x})
\right].$$
Since the standard product of operators
on a Hilbert space is an associative product, i.e.
$\hat A(\hat B \hat C)=(\hat A\hat B)\hat C$, 
it is obvious that the formula 
defines an associative product for the functions 
$f_{\hat A}({\bf x})$, namely,
$$f_{\hat A}({\bf x})*\Big(f_{\hat B}({\bf x})
*f_{\hat C}({\bf x})\Big)=
\Big(f_{\hat A}({\bf x})*f_{\hat B}({\bf x})\Big)
*f_{\hat C}({\bf x}).$$
The commutator of two operators 
$$\hat C=[\hat A, \hat B]=\hat A\hat B-\hat B\hat A $$
is mapped onto the quantum Poisson bracket$\,$\cite{Dirac}
 $f_{\hat C}({\bf x})$ of two
symbols $f_{\hat A}({\bf x})$ and $f_{\hat B}({\bf x})$ 
$$f_{\hat C}({\bf x})=\Big\{f_{\hat A}({\bf x}),
f_{\hat B}({\bf x})\Big\}_*=
\mbox{Tr}\left[[\hat A,\hat B]\hat U({\bf x})\right].$$
Since the Jacobi identity is fulfilled for the
commutator of the operators, 
the Jacobi identity is also fulfilled for the
Poisson bracket of the functions $f_{\hat A}({\bf x})$ and
$f_{\hat B}({\bf x})$. 
Since for operators one has the derivation property
 with respect
to the associative product,
the Poisson brackets reproduce this property
$$\Big\{f_{\hat A}({\bf x}),f_{\hat B}({\bf x})*
f_{\hat C}({\bf x})\Big\}_*=\Big\{f_{\hat A}({\bf x}),
f_{\hat B}({\bf x})\Big\}_**f_{\hat C}({\bf x})+
f_{\hat B}({\bf x})*\Big\{f_{\hat A}({\bf x}),
f_{\hat C}({\bf x})\Big\}_*
$$
this property  qualifies it as a {\it quantum Poisson bracket}.

 The evolution in the space of observables
can be described by the Heisenberg equations of 
motion  
$$\dot{\hat A}=i[\hat H,\hat A]\qquad (\hbar=1),$$
where $\hat H$ is the Hamiltonian of the system
and $\hat A$ is a generic observable.
This equation can be rewritten in terms of the functions
$f_{\hat A}({\bf x})$ and $f_{\hat H}({\bf x})$, where 
$$f_{\hat H}({\bf x})=\mbox{Tr}\left[\hat H\hat U({\bf x})\right]$$
corresponds to the Hamiltonian, in the form 
$$\dot f_{\hat A}({\bf x},t)=i\Big\{f_{\hat H}({\bf x},t),
f_{\hat A}({\bf x},t)\Big\}_*$$
via the quantum Poisson bracket we have defined.
 
Let us suppose that there exist other maps,
we choose two different ones.
One map is described by a vector ${\bf x}=(x_1,x_2,\ldots,x_n)$
and operators $\hat U({\bf x})$
and $\hat D({\bf x})$. 
Another map is described by a vector
${\bf y}=(y_1,y_2,\ldots,y_m)$ and operators
$\hat U_1({\bf y})$ and
$\hat D_1({\bf y})$. 
For given operator $\hat A$, one has 
the function
$$\phi_{\hat A}({\bf y})=\mbox{Tr}
\left[\hat A\hat U_1({\bf y})\right]$$
and the inverse relation
$$\hat A=\int\phi_A({\bf y})\hat D_1({\bf y})~d{\bf y}. $$
One can obtain a relation of the
function $f_{\hat A}({\bf x})$ to the function 
$\phi_{\hat A}({\bf y})$ in the form
$$\phi_{\hat A}({\bf y})=\int f_{\hat A}({\bf x})\,\mbox{Tr}\left[
\hat D({\bf x})\hat U_1({\bf y})\right]\,d{ \bf x}$$
and the inverse relation 
$$f_{\hat A}({\bf x})=
\int\phi_{\hat A}({\bf y})\,\mbox{Tr}\left[\hat D_1({\bf y})
\hat U({\bf x})\right]\,d{\bf y}.$$
We see that functions $f_{\hat A}({\bf x})$ and
$\phi_{\hat A}({\bf y})$ corresponding to different
maps are connected
by means of the invertible integral transform. 
These transforms are determined by means of intertwining 
kernels 
$$K_1({\bf x},{\bf y})=\mbox{Tr}\,\Big[\hat D({\bf x})
\hat U_1({\bf y})\Big ]\qquad\mbox{and}\qquad 
K_2({\bf x},{\bf y})=\mbox{Tr}\,\Big[\hat D_1({\bf y})
\hat U({\bf x})\Big].$$
One can write
down a composition rule for two symbols $f_{\hat A}({\bf x})$
and $f_{\hat B}({\bf x})$, which determines the star-product
of these symbols. The composition rule is described by the
formula 
$$f_{\hat A}({\bf x})*f_{\hat B}({\bf x})=
\int f_{\hat A}({\bf x}'')f_{\hat B}({\bf x}')
K\left({\bf x}'',{\bf x}',{\bf x}\right)d{\bf x}'\,d{\bf x}''.$$
The kernel under the integral is
determined by the trace of product of the basic operators,
which we use to construct the map 
$$K({\bf x}'',{\bf x}',{\bf x})=
\mbox{Tr}\left[\hat D({\bf x}'')\hat D({\bf x}')
\hat U({\bf x})\right].$$
 
\section{Symplectic tomograms} 

According to the general scheme one can introduce for the
operator $\hat A$ the function $f_{\hat A}({\bf x})$, where
${\bf x}=(x_1,x_2,x_3)\equiv (X,\mu,\nu)$,
which we denote here as
$w_{\hat A}(X,\mu,\nu)$ depending on the position $X$ and
the parameters $\mu$ and $\nu$
 of the reference frame
$$w_{\hat A}(X,\mu,\nu)=\mbox{Tr}\left[\hat A
\hat U({\bf x})\right].$$
We call the function $w_{\hat A}(X,\mu,\nu)$ the tomographic 
symbol of the operator $\hat A$. The operator $\hat U({\bf x})$ is 
given by
$$
\hat U({\bf x})\equiv \hat U(X,\mu,\nu)=
\delta\left(X-\mu\hat q-\nu \hat p\right),$$
where $\hat q$ and $\hat p$ are position and momentum 
operators.

In the case under consideration, the inverse transform 
will be of the form
$$\hat A=\int w_{\hat A}(X,\mu,\nu)
\hat D(X,\mu,\nu)\,dX\,d\mu\,d\nu,$$
where$\,$\cite{Dariano}
$$\hat D({\bf x})\equiv\hat D(X,\mu,\nu)=\frac{1}{2\pi}
\exp\left(iX-i\nu\hat p-i\mu\hat q\right).$$
If one takes two operators $\hat A_1$ and $\hat A_2$,
which are expressed through the corresponding functions
by the formulae
$$\hat A_1
=
\int
w_{\hat A_1}(X',\mu',\nu')\hat D(X',\mu',\nu')\,dX'\,d\mu'
\,d\nu'\,,$$
$$\hat A_2
=
\int
w_{\hat A_2}(X'',\mu'',\nu'')
\hat D(X'',\mu'',\nu'')dX''\,d\mu''\,d\nu''\,,$$
and $\hat A$ denotes the product of $\hat A_1$
and $\hat A_2$, then
the function $w_{\hat A}(X,\mu,\nu)$, which 
correspond
s to $\hat A$, is the star-product of functions 
$w_{\hat A_1}(X,\mu,\nu)$
and $w_{\hat A_2}(X,\mu,\nu)$, i.e.,
$$
w_{\hat A}(X,\mu,\nu)=w_{\hat A_1}(X,\mu,\nu)
*w_{\hat A_2}(X,\mu,\nu)
$$
reads
$$w_{\hat A}(X,\mu,\nu)=\int w_{\hat A_1}({\bf x}'')
w_{\hat A_2}({\bf x}')K({\bf x}'',{\bf x}',
{\bf x})\,d{\bf x''}\,d{\bf x'},$$
with the kernel given by
$$K({\bf x}'',{\bf x}',{\bf x})=
\mbox{Tr}\left[\hat D(X'',\mu'',\nu'')
\hat D(X',\mu',\nu')\hat U(X,\mu,\nu)\right].$$
The explicit form of the kernel reads
\begin{eqnarray*}
&&K(X_1,\mu_1,\nu_1,X_2,\mu_2,\nu_2,X\mu,\nu)=
\frac{\delta\Big(\mu(\nu_1+\nu_2)-\nu(\mu_1+\mu_2)\Big)}{4\pi^2}
\nonumber\\
&&\times
\exp\left[\frac{i}{2}\Big(\left(\nu_1\mu_2-\nu_2\mu_1\right)
+2X_1+2X_2
-\frac{2\left(\nu_1+\nu_2\right)}{\nu}\,X\Big)\right].\nonumber
\end{eqnarray*}
By using the oscillator's ground state
$$w_0=\left[\pi\left(\mu^2+\nu^2\right)\right]^{-1/2}
\exp\left(-\frac{X^2}{\mu^2+\nu^2}\right),$$
one can check that this kernel satisfies the obvious relation
$$w_0*w_0=w_0.$$

\section{Spin tomograms}

In order to construct spin-tomogram according to the 
general considerations, let us introduce the operator $U$
of finite 
rotation of the SU(2) group
.

Following~\cite{dod14,JETF}
we will derive the expression for an
arbitrary operator acting on spin states in terms of measurable mean
values of the operator for the spin projection on a given direction,
considered in a rotated reference frame.
For arbitrary values of spin, let the operator $\widehat A^{(j)}$
be represented by the matrix
$$A_{mm'}^{(j)}=\langle jm\mid \widehat A^{(j)}\mid jm'\rangle \,,
\qquad m=-j,-j+1,\ldots,j-1,j\,,$$
where
$$\hat j_3 \mid j m \rangle
=
m \mid j m \rangle\,;\qquad
\hat j^2 \mid j m\rangle
=
 j(j+1) \mid j m\rangle\,,$$
and
$$\widehat A^{(j)}=\sum_{m=-j}^j \sum_{m'=-j}^j A_{m
m'}^{(j)}\mid j m\rangle\langle j m'\mid \,;\qquad
A_{m m }^{(j)} = w_0(m)\,.$$
Let us condsider the
diagonal elements of the operator in another reference frame 
$$A^{(j)}_{m_1\,m_1}=\left ( U A U^\dagger
\right )_{m_1\,m_1}.$$
The unitary rotation transform $U$ depends on the Euler angles
$\alpha ,\,\beta ,\,\gamma \,$.

Below we introduce new notation for the diagonal
matrix elements of the operator
and rewrite the above equality in the form
$$\widetilde w\left(m_1,\alpha,\beta\right) =
\sum^j_{m_1'=-j}\,\sum^j_{m_2'=-j}
\,D_{m_1m_1'}^{(j)}(\alpha, \beta,\gamma)\,
A^{(j)}_{m_1'm_2'}\,D_{m_1m_2'}^{(j)\ast}(\alpha, \beta, \gamma)\,.$$
The defined function is called 
the tomographic symbol of the operator.
In view of the structure of the above formula,
the tomogram depends only on two Euler angles.
Here the matrix elements
$D_{m_1\,m_1'}^{(j)}\left (\alpha, \beta, \gamma \right)$~are the
matrix elements of the rotation transform $U$
$$D^{(j)}_{m'm}(\alpha,\beta,\gamma)=e^{i m'\gamma}\,d_{m'm}^{(j)}(\beta)\,
e^{i m \alpha}\,,$$
where
\begin{eqnarray*}
d_{m'm}^{(j)}(\beta)&=& 
\left[\frac{(j+m')!(j-m')!}{(j+m)!(j-m)!}\right]^{1/2}
\left(\cos\,\frac{\beta}{2}\right)^{m'+m} \left(\sin\,
\frac{\beta}{2}\right)^{m'-m}\nonumber\\
&&\times P_{j-m'}^{(m'-m,m'+m)}(\cos\,\beta)
\nonumber
\end{eqnarray*}
and $P_n^{(a,b)}(x)$ is the Jacobi polynomial.

The inverse relation reads
\cite{22klimov}
$$\sum_{j_3=0}^{2j}\,\sum_{m_3=-j_3}^{j_3}\,(2j_3+1)^2\sum_{m_1=-j}^{j}
\int (-1)^{m_1}\,\widetilde w\left(m_1,\alpha,\beta\right)\,D_{0\,m_3}^{(j_3)}
\left(\alpha,\beta,\gamma\right)$$
$$\otimes
\pmatrix{j&j&j_3\cr
m_1&-m_1&0}\,\pmatrix{j&j&j_3\cr m_1'&-m_2'&m_3}\,
\frac{d\omega}{8\,\pi^2}=(-1)^{m_2'}\,A_{m_1'm_2'}^{(j)},$$
where $d\omega$ is the volume element in terms
of Euler angles.
One can obtain an invariant form of this relation.
To do this, let us
introduce the function on the sphere
$$\Phi_{j m_1'm_2'}^{(j_3)}\left(\alpha, \beta \right )
=(-1)^{m_2'}\sum_{m_3=-j_3}^{j_3}
D_{0\,m_3}^{(j_3)}\left(\alpha,\beta,\gamma\right)
\pmatrix{j&j&j_3 \cr m_1'&-m_2'&m_3}$$
and the operator on the sphere
$$\widehat A_j^{(j_3)}\left(\alpha,\beta\right)
=(2j_3+1)^2\sum_{m_1'=-j}^{j}\,\sum_{m_2'=-j}^j
\mid j m_1'\rangle \,\Phi_{j m_1'm_2'}^{(j_3)}\left(\alpha,\beta\right)
\langle j m_2'\mid.$$
Then using another operator 
${\bf x}=\left(m_1,\alpha,\beta\right)$
$$\widehat B^{(j)}\left({\bf x}\right)\equiv
\widehat B_{m_1}^{(j)}\left(\alpha,\beta\right)
=(-1)^{m_1}\sum_{j_3=0}^{2j}\pmatrix{j&j&j_3
\cr m_1&-m_1&0} \widehat A_j^{(j_3)}\left(\alpha,\beta\right)\,,$$
one can write for the operator under study the invariant 
expression:
\cite{JETF}
$$\widehat A^{(j)}=\sum_{m_1=-j}^j
\int\frac{d\omega}{8\,\pi^2}\,\widetilde w 
\left(m_1,\alpha,\beta\right)\,\widehat
B_{m_1}^{(j)}\left(\alpha,\beta\right)\,.$$

Let us consider the tomographic star-product of two functions
$W(m_1,\alpha,\beta)$ and $W'(m_1,\alpha,\beta)$. 
One has, by definition
,
$$\left(W\star W'\right)\left(m_1,\alpha,
\beta\right)=\sum_{{\bf x}',{\bf x}''}
 W \left(m'_1,\alpha',\beta'\right)
W'\left(m''_2,\alpha'',\beta''\right)
K({\bf x},\bf{x}',{\bf x}''),$$
where the kernel has the form 
$$K({\bf x},{\bf x}',{\bf x}'')=\frac{1}{\left(8\,\pi^2\right)^2}
\mbox{Tr}\left[|jm_1\rangle\langle j m_1|
U\,\widehat B^{(j)}({\bf x}')\,
\widehat B^{(j)}({\bf x}'')\,
U^\dagger\right].$$
The kernel can be given in explicit form. The tomographic star-product
defined by means of the above kernel defines an associative product.
So we have an invertible map between the operator $\widehat A^{(j)}$ and
its symbol $\widetilde w({\bf x})$
:
$$\widehat A^{(j)}
=
\sum_{\bf x}\widetilde w({\bf x})\widehat D({\bf x}),\qquad
\widehat D({\bf x})=\frac{1}{8\pi^2}\widehat B^{(j)}_{m_1}({\bf x}),$$
$$\widetilde w({\bf x})
=
\mbox{Tr}[\widehat A^{(j)}\widehat{\tilde U}
({\bf x})],\qquad\widehat{\tilde U}({\bf x})=
U^\dagger|j m_1\rangle\langle j m_1|U\,,$$
where sum over ${\bf x}$ means the integral over Euler 
angles and sum over angular momentum projection.

\section{Conclusions}

To conclude, we have embedded the tomographic map
into the general scheme of symbols and operators.
When the kernel of star-product is known,
Heisenberg equations of motion for observables 
written in the form of equations for their symbols 
can be presented, in some cases, in the form of partial 
differential equations. For example, in the tomographic 
representation, the evolution equation for density matrix
takes the form of Fokker--Planck-type equation.\cite{Tombesi13}
One can also obtain such differential form of the evolution
equation for spin tomograms.

\section*{Acknowledgments}

Olga~V.~M. and Vladimir~I.~M. thank the organizers for kind
hospitality, the Russian 
Foundation for Basic Research for partial support 
under Projects~Nos.
99-02-17753, 
00-02-16516, and 01-02-17745
and the Ministry for Industry, Sciences and Technology of 
the Russian Federation for the support within the framework
of the Programs ``Optics. Laser Physics'' and ``Fundamental
Nuclear Physics.'' 
Vladimir~I.~Man'ko is grateful to the 
Russian Foundation for Basic Research for Travel Grant
No.~01-02-26874.


\section*{References}
\begin{thebibliography}{99}

\bibitem{[1]} F.~Bayen, M.~Flato, C.~Fronsdal, A.~Lechnerovicz, 
and D.~Sternheimer,
 {\it Lett. Math. Phys.}, {\bf 1},  521
(1975).

\bibitem{Tombesi13} S.~Mancini, V.~I.~Man'ko, and P.~Tombesi,
{\it Phys. Lett. } A, {\bf 213}, 1 (1996).

\bibitem{Tombesi13'} S.~Mancini, V.~I.~Man'ko, and P.~Tombesi,
{\it Found. Phys.}, {\bf 27}, 801 (1997).

\bibitem{dod14} V.~V.~Dodonov and V.~I.~Man'ko,
{\it Phys. Lett.}, {\bf 239}, 335 (1997).

\bibitem{JETF} V.~I.~Man'ko and O.~V.~Man'ko,
{\it JETP}, {\bf 85}, 430 (1997).

\bibitem{21klimov} V.~I.~Man'ko and G.~Marmo,
{\it Phys. Scr.}, {\bf 58}, 224 (1998).

\bibitem{21'klimov} V.~I.~Man'ko and G.~Marmo,
{\it Phys.~Scr.}, {\bf 60}, 111 (1999).

\bibitem{22klimov} O.~V.~Man'ko, V.~I.~Man'ko, and G.~Marmo,
{\it Phys. Scr.}, {\bf 62}, 446 (2000).

\bibitem{JPAmarmotobepublished} 
O.~V.~Man'ko, V.~I.~Man'ko, and G.~Marmo,
{\it J. Phys.} A (submitted).

\bibitem{Dirac}
P. A. M. Dirac, {\it The Principles of Quantum Mechanics},
4th ed., Pergamon, Oxford (1958).

\bibitem{Dariano}
 G.~M.~D'Ariano, S.~Mancini, V.~I.~Man'ko, and P.~Tombesi, 
{\it J. Opt.} B, {\bf 8}, 1017 (1996).

\end{thebibliography}
\end{document}